\documentclass[preprint]{elsarticle}
\usepackage{lineno,hyperref}
\modulolinenumbers[5]
\journal{Journal of \LaTeX\ Templates}

\usepackage{dcolumn}
\usepackage[utf8]{inputenc}
\usepackage[english]{babel}
\usepackage{amsmath}
\usepackage{graphics}
\usepackage{verbatim}
\usepackage{amssymb}
\usepackage{array}
\usepackage{listings}
\usepackage{hyperref}
\usepackage{bookmark}
\usepackage{tabularx}

\begin{document}

\begin{frontmatter}

\title{Atmospheric density uncertainty effects on the orbital lifetime estimation for CubeSats at LEO}

\author[]{$^{*}$D. J. Cubillos Jara$^{1,2,3}$}
\ead{dcubillosj@ecci.edu.co}
\author[]{J. A. Soliz Torrico$^3$}
\ead{soliz.jorge@gmail.com}
\author[]{O. L. Ramírez Suárez$^{1,4}$}
\ead[]{oramirezs@ecci.edu.co}
\address{$^1$Grupo de Simulación, Análisis y Modelado en Ciencias Básicas (SiAMo), Universidad ECCI, Bogotá, Colombia.}
\address{$^2$ Grupo de Instrumentación Radio Astronómica (R.A.I.G), Facultad de Ciencias Físicas y Matemáticas, Departamento de Ingeniería Eléctrica, Universidad de Chile, Santiago, Chile.}
\address{$^3$Departamento de Ciencias Básicas, Universidad ECCI, Bogotá, Colombia.}
\address{$^4$Ciencias exactas, Facultad de Ingeniería, Universidad Privada Boliviana, Cochabamba, Bolivia.}
\address{$^5$Vicerrectoría de Investigación, Universidad ECCI, Bogotá, Colombia.}




\begin{abstract}
Nanosatellites, and especially CubeSats, at low earth orbits (LEOs) are a low cost option for monitoring atmospheric and environmental conditions around Earth. For instance, data for weather forecast reports can be obtained periodically with these kind of small satellites. Therefore, to academic institutions, universities, etc., this fact makes nanosatellites a very attractive way for researching with a moderate budget.

In this project we compute orbital lifetimes (or simply lifetimes) for hypothetical missions involving nanosatellites at LEO, focusing our attention on exploring regions along the equatorial line. Thus, in the framework of orbital mechanics, we show the viability for these kind of missions in a long and a short term. Applications are projected for countries in northern South America, central Africa and islands/countries in southern Asia. 

To compute lifetimes, we take into account three effects: i) gravitational, ii) Earth deformations and iii) atmospheric density. These effects are included in the motion equation for a nanosatellite around Earth. After solving this equation for initial altitudes in 200-800 km above mean sea level (AMSL), we compute and report flight times to arrive at 150 km AMSL. These results are defined here as lifetimes and they are calculated for different atmospheric-density profiles according to experimental data and estimations.

In conclusion, we find lowest and highest lifetimes for hypothetical missions involving small satellites at LEO orbiting along the equatorial line, and propose upper limits for density relative uncertainties in order to estimate reliable lifetimes.
\end{abstract}


\end{frontmatter}

%\linenumbers

\section{Introduction}

Estimations of orbital lifetimes from initial conditions and the dynamics of a satellite have been a challenge for any space mission \cite{Sharaf2013,ElSalam} (and references therein). Nowadays, lifetime estimations are more and more necessary and required due to the huge improvement and development of the so-called CubeSats (see e.g.\ Ref.\ \cite{Diaz} and references therein). These small satellites have opened the spectrum of possibilities for exploring our planet. For instance, academic missions of low cost such as Munin, Astrid and Astrid-2 \cite{Marklund} have collected data of  auroral activity, electric and magnetic fields in the upper ionosphere and neutral, charged particles density and, of course, telemetry.

With the increasing amount of CubeSat missions, which usually operate few years because of the internal electronics, more debris are expected in a short term and they can compromise the success of new missions \cite{Walker}. Therefore, if the operational lifetime can be increased to the order of the orbital lifetime, new missions will not need to be renewed avoiding an excess of debris in the future. On the other hand, by increasing operational lifetimes, all small missions can collect more information without any extra budget because no new mission, at least in the short term, will be required.

In this paper, we compute orbital lifetimes with the aim of: i) estimating the maximum interval of time which a CubeSat should operate by knowing initial orbital conditions only, ii) quantifying the effects of the atmospheric uncertainty on the orbital lifetime and iii) exploring the viability, in a short and a long term, of mission to monitor regions on the equatorial line.

The paper is organized as follows. In Sec.\ \ref{sec1} we discuss the dynamics of a CubeSat orbiting at LEO, we also classify the interactions and analyze the atmospheric density profile and its uncertainties. In Sec.\ \ref{sec2} the numerical procedure to estimate lifetimes is described. Results and the discussion are shown in Sec.\ \ref{sec3}. Conclusions and perspectives are summarized in Sec.\ \ref{sec4}.
\section{Interactions and motion equations for satellites at LEO}\label{sec1}
The motion of a small satellite,  specif-i-cally nanosatellite, orbiting around the Earth is explained via the Newton's second law of motion. In this case, the classification of all and most relevant interactions makes the problem much simpler or harder to solve.

The motion equation can be written as
\begin{equation}\label{EM}
m\vec{a}=\vec{F}_G+\vec{F}_D+\vec{F}_{TB}+\vec{F}_R+\cdots,
\end{equation}
where we assume that the satellite neither loses nor gains mass, and four of the most relevant interactions: gravitational (satellite - Earth), drag, third body and radiation are denoted by the subscripts $G$, $D$, $TB$ and $R$ respectively. Any other interaction can be included as indicated in Eq.\ \eqref{EM}.

Both interactions, $\vec{F}_{TB}$ and $\vec{F}_R$, are negligible in comparison to $\vec{F}_G$ and $\vec{F}_D$, specially at low altitudes. Therefore, in order to estimate lifetimes we shall consider $\vec{F}_G$ and $\vec{F}_D$ contributions only. It is clear that we cannot neglect $\vec{F}_D$ because this term is responsible of reducing the total energy of the satellite; otherwise the lifetime becomes infinity.

Before discussing $\vec{F}_G$ and $\vec{F}_D$ in detail, let us introduce and organize, in Table \ref{Tparameters}, all relevant constants and parameters to compute lifetimes according to Eq.\ \eqref{EM} and our assumptions.

\begin{table}[htb]
	\caption{Parameters and notation. $G$, $M$ and $R$ are taken from Ref.\ \cite{PDG}, $J_2$, $J_4$, etc., are taken from Ref.\ \cite{Kozai} and $C_D$ is taken as the standard value \cite{Gaposchkin}.}
	\label{Tparameters}
	%\begin{ruledtabular}
		\begin{tabular}{lcc}\hline\hline
			Parameter					& Symbol	& Value or  range \\
			\hline
			Gravitational constant		& $G$	 	& $6.67384\times10^{-11}$ m$^3/$kg/s$^2$\\
			Earth mass					& $M$ 		& $5.9722\times10^{24}$ kg\\
			Earth radius				& $R$ 		& 6371 km \\
			Satellite mass				& $m$ 		& 1 kg - 10 kg\\
			Satellite altitude			& $h$ 		& 150 km - 800 km \\
			Initial satellite altitude	& $h_0$ 	& 200 km - 800 km \\
			Initial orbital radius		& $R_0$		& $R+h_0$ \\
			Initial speed				& $v_0$		& $\sqrt{GM/R_0}$ \\
			Coefficient J2				& $J_2$		& $1.082645(6)\times10^{-3}$ \\
			Coefficient J4				& $J_4$		& $-1.649(16)\times10^{-6}$ \\
			Coefficient J6				& $J_6$		& $0.646(30)\times10^{-6}$ \\
			Coefficient J8				& $J_8$		& $-0.270(50)\times10^{-6}$ \\
			Coefficient J10				& $J_{10}$	& $-0.054(50)\times10^{-6}$ \\
			Coefficient J12				& $J_{12}$	& $-0.357(44)\times10^{-6}$ \\
			Coefficient J14				& $J_{14}$	& $0.179(63)\times10^{-6}$ \\
			Drag coefficient 			& $C_D$		& 2.2 \\ \hline
		\end{tabular}%\hline
	%\end{ruledtabular}
\end{table}

\subsection{Gravitational interaction}
The most relevant interaction to study an orbital motion is given by the Newton's law of universal gravitation. In the simplest case, where two point objects interact gravitationally, this interaction can be written as
\begin{equation}\label{UGL}
	\vec{F}=-G\frac{M_T\mu}{r^3}\vec{r},
\end{equation}
with $G$ the universal gravitational constant or Newton's constant, $M_T$ and $\mu$ the total mass and the reduced mass of the bodies, $\vec{r}$ the relative position of $\mu$ with respect to $M_T$ and $r$ the magnitude of $\vec{r}$. It is important to note that, although we shall study a two-body problem, the huge difference between the mass of the Earth and the mass of the satellite makes the total mass and the reduced mass correspond approximately to $M$ and $m$ (see Table \ref{Tparameters}) respectively. Therefore, in the center of mass frame we can neglect the motion of the Earth and consider the motion of the satellite only.

On the other hand, the Earth should not be considered as a point or spherical object. The Earth shows an oblateness along its rotational axis, and also, it is not symmetric rotationally. In order to consider these deformations, the gravitational interaction can be generalized via the so-called $J_n$ terms.

The generalization of the gravitational potential $U$ for a spheroid of revolution is 
\begin{equation}\label{UGL2}
\vec{F}=-\vec{\nabla}U,
\end{equation}
where $U$ reads \cite{Chobotov}
\begin{equation}\label{pot}
U=-\frac{GMm}{r}\left[1-\sum_{n=2}^\infty\left(\frac{R}{r}\right)^nJ_nP_n(w)\right],
\end{equation}
$J_2$, $J_4$, $J_6$, etc., are displayed in Table \ref{Tparameters} (see Ref.\ \cite{Chobotov} for more details about low order coefficients, or see Ref.\ \cite{Kozai} for coefficients up to $J_{14}$), $P_n(w)$ is the Legendre polynomial of order $n$, and $w=\sin(\delta)$ with $\delta$ the declination of satellite.

Let us now discuss briefly the $J_n$ term and the $J_n$ contribution (i.e., $J_n$ term together with all its multiplicative factors in Eqs.\  \eqref{UGL2} and \eqref{pot}) for the Earth as follows.
\subsubsection{J{\tiny{2}} contribution}%$J_2$
This term describes the oblateness of the Earth and, in contrast with other $J_n$ terms, it is expected to be the highest contribution\cite{Chobotov}.

According to Eqs.\ \eqref{UGL2} and \eqref{pot}, the farther the satellite is from the Earth, the weaker is the $J_n$ contribution. In particular, the $J_2$ contribution is less than a thousand times weaker than the first term on the r.h.s.\ of Eq.\ \eqref{UGL2} (after replacing Eq.\ \eqref{pot}) for  satellites orbiting in the region 200 to 800 km ASML (i.e., at LEO).

\subsubsection{J{\tiny{2n+1}} contribution for n greater than zero}%$J_{2n+1}$ $n>0$
\begin{figure}[htb]
	\includegraphics[width=0.90\textwidth,height=0.30\textheight]{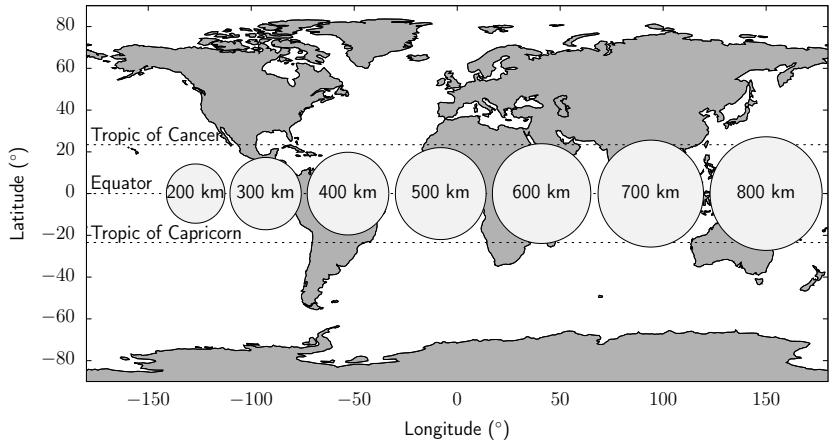}%width=7.6cm
	\caption{Observed Earth surface by a nanosatellite orbiting on the equatorial plane at altitude from 200 km to 800 km AMSL. The altitude is given in the center of each circle.}
	\label{area}
\end{figure}

In this study we focus our attention on those satellites orbiting on the equatorial plane. This fact makes the calculation much simpler because $\delta$ in Eqs.\ \eqref{UGL2} and \eqref{pot} vanishes. Therefore, any $J_{2n+1}$ contributions disappear for $n>0$. This assumption is the basis for missions focused on monitoring the Earth between the tropics of Cancer and Capricorn ($\pm 23.44^\circ$ latitude). As Fig.\ \ref{area} shows, in this region we can cover from Bolivia and the northern of Australia up to the southern of Mexico and India. Moreover, along the equatorial line we find the following countries: Ecuador, Colombia, Brazil, Sao Tome \& Principe, Gabon, Republic of the Congo, Democratic Republic of the Congo, Uganda, Kenya, Somalia, Maldives, Indonesia and Kiribati. Note that the first three countries listed above are in South America.

Figure \ref{area} shows the area observed from the satellite as a function of the altitude. This can be crucial for determining the initial conditions for a mission, and also illustrates the relevance of our study for monitoring countries along the equatorial line.

It is important to stress that other kind of orbits, different to the equatorial ones, can be used to monitor these countries. However, the revisiting time can be longer if no propulsion system is included.

\subsubsection{J{\tiny{2n+2}}  contribution  for n greater than zero}%$J_{2n+2}$$n>0$
Compared to the $J_2$ contribution, all the $J_\text{even}$ contributions are a factor of a thousand or more lower which can be inferred from Table \ref{Tparameters}. Moreover, this factor is magnified as the relative distance between the satellite and the Earth increases. However, we shall explore and show explicitly the contribution given by the first seven $J_\text{even}$ contributions on the lifetime estimation.

\subsection{Drag interaction}
Drag interaction is the most important force that reduces the orbital lifetime for a satellite without a propulsion system. This force is given  by
\begin{equation}\label{drag}
\vec{F}_D=-\frac{1}{2}\rho v^2 C_D A \hat{v},
\end{equation}
where the subscript $D$ refers to drag, $\rho$ is the atmospheric mass density, $\vec{v}$ is the velocity of the satellite with respect to atmosphere ($v$ and $\hat{v}$ are the magnitude and unitary vector respectively), $C_D$ is known as the drag coefficient and $A$ is the satellite effective (projected) area. We shall consider that the orientation of the satellite is unchanged with respect to Earth and that $A$ corresponds to the area of the smallest face of the CubeSat (for simplicity we shall assume a CubeSat of size 1, 2 or 3 units or 1U, 2U or 3U, where the largest face is always pointing to Earth).

In this study we consider two cases: a static and a moving atmosphere, which means that the velocity, $\vec{v}$, is computed as $\vec{v}=\vec{v}_S-\vec{v}_A$, where the subscripts $S$ and $A$ stand by satellite and atmosphere respectively. Both velocities ($\vec{v}_S$ and $\vec{v}_A$) are measured in the geocentric frame. In the static case $\vec{v}_A$ vanishes.

On the other hand, an accurate atmospheric profile is desirable as much as possible. This term affects the equation of motion directly and locally. Moreover, $\rho$ is not straight forward to determine experimentally. Thus, we dedicate the next section to discuss how we assume such a profile.

\subsubsection{Atmospheric density approach for altitudes in 150 - 800 km AMSL}

As we shall not consider any solar effect, it makes sense that the atmosphere is rotationally symmetric in the range 150-800 km ASML. In a more detailed study, day and night effects on the atmosphere can in principle be included.

Here, we shall assume that the mass density follows the profile reported in Table I of Ref.\ \cite{NASA}. We wish to stress that no oblateness, nor any deformation, of the atmosphere are taken into account because we analyze only orbits on the equatorial plane.

\begin{figure}[htb]
	\includegraphics[width=0.90\textwidth,height=0.30\textheight]{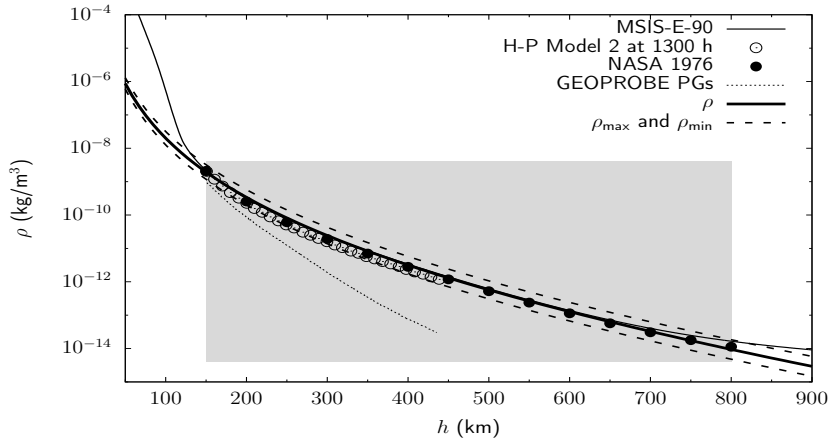}%[width=7.6cm]
	\caption{Atmospheric density profile according to Refs.\ \cite{Pelz,MSISE90,NASA} and our parametrization ($\rho$) with its highest and lowest limits ($\rho_\text{max}$ and $\rho_\text{min}$). The mathematical function for the parametrization and its limits are shown in the text. The shaded region highlights the range where the parametrization is used. Data sets are taken from: Ref.\ \cite{MSISE90} to MSIS-E-90, Ref.\ \cite{Pelz} to H-P Model 2 at 1300 h and GEOPROBE PGS, and Ref.\ \cite{NASA} to NASA 1976. Although the data set from NASA 1976 has significantly more points, we plot only those points between 150 and 800 km in steps of 50 km for the sake of clarity.}
	\label{atm}
\end{figure}

Figure \ref{atm} shows the mass density profile according to Refs.\ \cite{MSISE90,Pelz,NASA} (see also Ref.\ \cite{Rocken}). 

In order to optimize the lifetime computation, we have parametrized the data from Ref.\ \cite{NASA} as
\begin{equation}\label{densi}
\rho=\frac{a}{h^d}\exp\left(-\frac{h}{b+c\,h}\right),
\end{equation}
where $a=336.92$ kg\,m$^2$, $b=161.30$ km, $c=0.0071$, $d=5$ and $h$ is the satellite altitude.

With the previous parametrization we do not attempt to provide a density model. We just propose a mathematical function capable of reproducing in a good approximation the data in Ref.\ \cite{NASA} in the range 150-800 km AMSL.

Similarly, we propose a maximum and a minimum density ($\rho_\text{max}$ and $\rho_\text{min}$) with the same mathematical structure that Eq.\ \eqref{densi} but changing $c\to 0.98c$ and $d\to 1.02d$ for $\rho_\text{max}$, and $c\to 1.02c$ and $d\to 0.98d$ for $\rho_\text{min}$. 

For high altitudes ($h>86$ km \cite{NASA}) densities are not measured directly and then models need to be proposed. Thus, we shall use the three densities, $\rho_\text{max}$, $\rho$ and $\rho_\text{min}$, which from now on will be named as highest, mean and lowest density, to simulate uncertainties in the atmospheric density data set reported in Ref.\ \cite{NASA}, in order to see the effects on lifetime estimations.

\section{Numerical considerations}\label{sec2}
We solve Eq.\ \eqref{EM} by decomposing it in a Cartesian coordinates system, where the Z axis goes from south pole to north pole and the X axis is always chosen such that the initial position of the satellite is written as $\vec{r}_0=r\hat{i}$. Among the three coupled equations that for de $z$ component is trivially solved due to we are assuming orbits on the equatorial plane.

We develop a code in FORTRAN language which solves the system of equations via the Runge-Kutta-Fehlberg method \cite{RKF} of orders 7 and 8. The initial conditions are assumed as an circular orbit assuming the Earth as a point object (for initial conditions only). Table \ref{Tparameters} displays the range of initial orbital radius and initial speed. 

Here, the lifetime is defined as the total interval of time that the satellite spends from the initial conditions up to when the satellite arrives at 150 km AMSL.

In order to check the stability of the numerical integration, we perform two calculations. First, we define and compute $\Delta=|h_0-h_{100}|$, where $h_0$ is already defined in Table \ref{Tparameters} and $h_{100}$ is the altitude AMSL after a century of orbiting around the Earth. In this case we turn off all the interactions except the gravitational one by assuming the Earth as a point mass and considering circular orbits only. For $h_0$ in 200-800 km the computation gives us $\Delta< 2$ cm, which shows that numerical errors will be negligible for our lifetime estimations. The second calculation to test the numerical stability of the code, consists on changing the Runge-Kutta-Fehlberg stepsize up/down to find numerical convergence. If the numerical tolerance is not satisfied, the numerical integrator is capable of redefining the stepsize in a predefined interval (adaptive stepsize). We keep the maximum and minimum of this interval between 1 and $10^{-5}$, and the stepsize is setup initially as $10^{-n}$ with $n\in\{1,2,3,4\}$. In this test, we allow the Earth to be a deformed object and we include atmospheric effects. Assuming a CubeSat of 1U of 0.5 kg or 1 kg and $h_0$ in the range 200-800 km, the second test provides a lifetime uncertainty lower than one part per thousand. Again, this numerical error is negligible. 

\section{Results and discussion}\label{sec3}
\begin{figure}[htb]
	\includegraphics[width=0.90\textwidth,height=0.30\textheight]{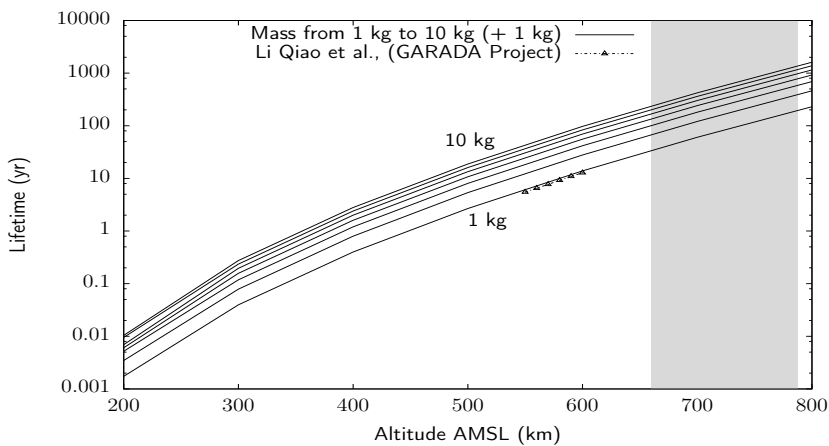}
	\caption{Lifetimes for hypothetical missions orbiting on the equatorial plane at LEO. No propulsion system is considered and the geometry and masses correspond to nanosatellites as it is described in the text. The shadow region highlights typical release altitudes. Lines from bottom (1 kg) to top (10 kg) differ sequentially in 1 kg. For a easier reading, the small tics on the lifetime axis represent a factor of ten $\times 2$, $\times 5$ and $\times 8$.}
	\label{life}
\end{figure}

Let us first discuss our more general results which are shown in Fig.\ \ref{life}. Here we compute lifetimes for CubeSats from 1 kg to 10 kg (e.g., CubeSats of 1U, 2U and 3U with equipment) on equatorial orbits (no inclination), nadir orientations and with all the first fourteen $J_n$ contributions turned on. An area, $A=100$ cm$^2$ (in Eq.\ \eqref{drag}) is assumed for all cases. The difference among the simulations is given by the mass as indicated in the figure.

In Fig.\ \ref{life}, we adopt $\vec{v}_A=\omega_E(\hat{k}\times\vec{r})$ from Ref.\ \cite{Lamberto} as our atmospheric velocity, where $\omega_E=2\pi$/day.

As expected, Fig.\ \ref{life} shows clearly the effect of the inertia, i.e., for higher masses, longer lifetimes. However, we observe that the lifetime ($\tau$) does not increase linearly with the mass. Nevertheless, the ratio $\tau_{(10\text{ kg})}/\tau_{(1\text{ kg})}$ is almost constant. For instance, rounding up to the second fractional number this ratio is 9.00, 9.90, 9.99, 10.00, 10.00, 10.00 and 10.00 for 200, 300, 400, 500, 600, 700 and 800 km AMSL respectively.

On the other hand, in Fig.\ \ref{life} we have also plotted the preliminary results of GARADA mission obtained by Li Qiao et al. \cite{Qiao} via the STK software \cite{STK}. As we can see the best agreement between Li Qiao et al.\ results and ours correspond to the line of 1 kg in Fig.\ \ref{life}. This shows two facts: first, the ratio $A/m$ is similar for both cases ($A/m=0.010$ m$^2$/kg according to our assumptions on a CubeSat of 1U and $A/m=0.013$ m$^2$/kg according to GARADA project), and second, the orientation of the orbit and other non-gravitational and atmospheric perturbations play a secondary role on the lifetime estimation. Note that the estimated lifetimes for the GARADA mission fall below our results, which is expected because $A/m$ is slightly higher for GARADA.

Another interesting result is how the $J_\text{even}$ contributions affects the lifetime estimation. We compute lifetimes progressively by activating, in the gravitational potential, all the terms up to $J_2$, then up to $J_4$, and so on up to $J_{14}$. Comparing the results we find that the lifetime decreases less than $0.5\%$ when more $J_\text{even}$ are included. The $J_\text{even}$ contributions play a secondary role because the density uncertainty can affect even more the lifetime estimation. This will be discussed in more detail below.

\begin{figure}[htb]
	\includegraphics[width=0.90\textwidth,height=0.30\textheight]{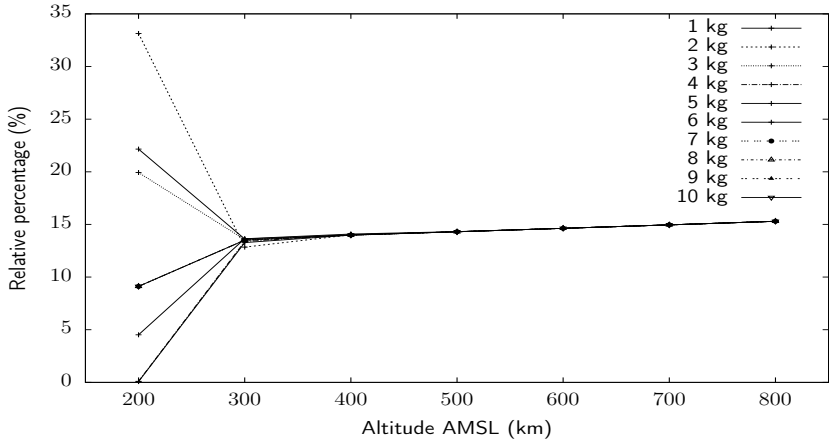}
	\caption{Relative percentage between lifetimes assuming a co-rotating and a static atmosphere. The percentage is computed as $(\tau_\text{co-rot}-\tau_\text{static})/\tau_\text{static}\times 100\%$. The longest lifetime correspond to the co-rotating case as it is explained in the text. Use Fig.\ \ref{life} as a reference point.}
	\label{life-sta}
\end{figure}

As mentioned above, we have also computed lifetimes for a static atmosphere. In Fig.\ \ref{life-sta} the relative percentage between lifetimes for a co-rotating atmosphere (Fig.\ \ref{life}) and lifetimes for a static atmosphere are shown. We plot this percentage instead of lifetimes as in Fig.\ \ref{life}, because we want to highlight the small differences between both lifetimes.

As one can see, for initial altitudes of 300 km AMSL or more, a co-rotating atmosphere makes lifetimes $\approx15$\% longer in comparison with those lifetimes assuming a static atmosphere. Note that the factor ($\approx15\%$) is almost constant and it is independent of the mass as Fig.\ \ref{life-sta} shows for $h\geq 300$ km AMSL. These results are expected because in our calculation we have assumed all satellites orbiting in the same direction as the motion of atmosphere (co-rotating case), which reduces the speed $v$ in Eq.\ \eqref{drag}.

Comparing the regions above and below 300 km AMSL in Fig.\ \ref{life-sta}, we see that both behaviors are clearly different. We think that this change is due to the relatively short trajectory and time (see Fig.\ \ref{life}) that the satellite takes to arrive to 150 km AMSL. However, a more detailed analysis should be performed in the range 200 - 300 km AMSL to provide strong conclusions.

Let us now analyze atmospheric-uncertainty effects on the lifetime estimation. As shown in Fig.\ \ref{atm} we have assumed that the uncertainty on the atmospheric density increases progressively with altitude. This assumption is justified because at higher altitudes the atmospheric density decreases, and therefore for a given instrument with constant resolution, the relative error for the density increases. This behavior on the atmospheric density is propagated to the lifetime estimation as expected. However, we note that the atmospheric relative uncertainty and the lifetime relative uncertainty are very similar (the latter slightly lower), as Table \ref{Ttau} shows.

\begin{figure}[htb]
	\includegraphics[width=0.90\textwidth,height=0.30\textheight]{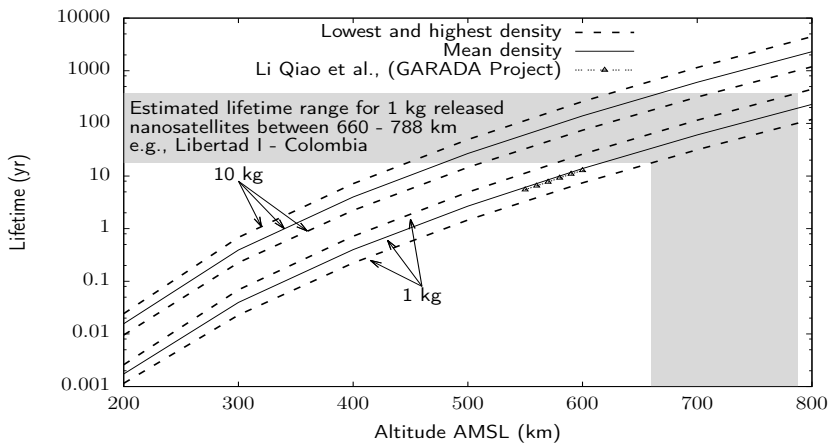}
	\caption{Same as Fig.\ \ref{life} but considering the effects of atmospheric uncertainties as shown in Fig.\ \ref{atm}. Only satellites of 1 kg and 10 kg are analyzed for the sake of clarity. Details about Libertad I can be found in Ref.\ \cite{Portilla}.}
	\label{life-unc}
\end{figure}

\begin{table}[htb]
	\caption{Relative lifetime estimations according to uncertainties given by the atmospheric density in Fig.\ \ref{atm}. The maximum, mean and minimum lifetimes ($\tau_\text{max}$, $\tau$ and $\tau_\text{min}$), are obtained by using the minimum, mean and maximum density profile ($\rho_\text{min}$, $\rho$, $\rho_\text{max}$). Values in the last two columns differ beyond the fourth fractional digit.}
	\label{Ttau}
	%\begin{ruledtabular}
		\begin{tabular}{cccccc}\hline\hline
			$h$ (km)	& $m$	(kg) & $\tau_\text{max}/\tau$	& $\tau/\tau_\text{min}$ & $\rho_\text{max}/\rho$	& $\rho/\rho_\text{min}$\\
			\hline
			200  		& 1		& 1.50	& 1.50 & 1.70 & 1.70\\
			200 		& 10 	& 1.56	& 1.64 & 1.70 & 1.70\\
			800			& 1		& 1.93	& 1.93 & 1.96 & 1.96\\
			800			& 10	& 1.93	& 1.93 & 1.96 & 1.96\\ \hline
		\end{tabular}
	%\end{ruledtabular}
\end{table}

\section{Conclusions and perspectives}\label{sec4}
We compute lifetimes for small satellites (CubeSat of from 1 to 10 kg of the type 1U, 2U or 3U) orbiting on the equatorial plane. Modifications to the spherical potential is taken into account due to the Earth deformations (up to the $J_{14}$ contribution only). We consider also the drag interaction because of the atmosphere and uncertainty effects on the density.

Effects of $J_\textbf{even}$ contributions on lifetime estimations play a secondary role in comparison with those given by an inaccurate density. By assuming the gravitational potential up to the $J_2$ contribution, and comparing the respective lifetime estimation with those obtained with more $J_\text{even}$ contributions, we see a change of less than $0.5\%$.

Results show a wide range of lifetimes when an inaccurate density profile is adopted. This effect is observed by simulating a highest- and a lowest- density profile. This density range is assumed approximately as a factor of two above and below the accepted density profile which has been estimated (not measured) at altitudes larger than 150 km AMSL. The density uncertainty is propagated to the lifetime in such a way that the lifetime relative uncertainty is similar but slightly lower than that for the density. This lead us to infer that, in order to decrease the relative lifetime uncertainty down to $10\%$, the atmospheric-density-profile uncertainty should not exceed the accepted profile by a factor of 1.2 approximately.

On the other hand, an inaccurate assumption for the co-rotating atmosphere effect can lead to wrong lifetime estimations in a factor of 15\% approximately. This effect can be significant for altitudes of 300 km AMSL or higher. For lower altitudes, we propose to perform a more detailed analysis to be able to come to similarly strong conclusions.

\section*{Acknowledgments}
We warmly thank SiAMo members for valuable discussions. We would like to thank G. Chaparro Molano for drawing our attention to Ref.\ \cite{Pelz} and for several suggestions. 

\section*{References}

%\bibliography{mybibfile}

\end{document}